# FACTORS INFLUENCING E-COMMERCE ADOPTION BY RETAILERS IN SAUDI ARABIA: A QUALITATIVE ANALYSIS

| Rayed AlGhamdi | Steve Drew | Waleed Al-Ghaith |
|---|---|---|
| rayed.alghamdi@griffithuni.edu.au | s.drew@griffith.edu.au | w.alghaith@griffith.edu.au |
| *School of Information and Communication Technology Griffith University, Brisbane, Australia* | *School of Information and Communication Technology Griffith University, Brisbane, Australia* | *Dept of Management, Griffith Business School Griffith University, Brisbane, Australia* |

**ABSTRACT**

This paper presents the preliminary findings of a study researching the diffusion and the adoption of online retailing in Saudi Arabia. It reports new research that identifies and explores the key issues that positively and negatively influence retailers in Saudi Arabia regarding the adoption of electronic commerce. Retailers in Saudi Arabia have been reserved in their adoption of electronically delivered aspects of their business. Despite the fact that Saudi Arabia has the largest and fastest growth of ICT marketplaces in the Arab region, e-commerce activities are not progressing at the same speed. Only a tiny number of Saudi commercial organizations, mostly medium and large companies from the manufacturing sector, are involved in e-commerce implementation. Based on qualitative data, collected by conducting interviews with a sample population of retail sector decision makers in Saudi Arabia, both positive and negative issues influencing retailer adoption of electronic retailing systems in Saudi Arabia are identified. A number of impediments which include cultural, business and technical issues were reported. Facilitating factors include access to educational programs and awareness building of e-commerce, government support and assistance for e-commerce, trustworthy and secure online payment options, developing strong ICT infrastructure, and provision of sample e-commerce software to trial. While literature reveals that government promotion has had limited effects on the diffusion of e-commerce in most countries, this study significantly indicates government promotion and support as a key driver to online retailing in KSA.

## 1.    INTRODUCTION

Comprehensive literature review reveals no investigation of the issues surrounding the development and diffusion of e-commerce in the context of KSA. This paper presents the preliminary findings of a study looking into the diffusion of the adoption of online retailing in Saudi Arabia. While the broader study includes all involved parties in retailing businesses (retailers, customers, and third parties including the government), this paper focuses on the retailers' perspectives. The paper reports findings that identify and explore the key issues that positively and negatively influence retailers in Saudi Arabia to adopt an online retail channel. The study uses a qualitative approach to explore the range of issues. In the next stage of research a quantitative approach will be used to test the findings in a wider sample.

Many businesses around the world have introduced an electronic commerce (e-commerce) channel as part of their operations, seeking the many advantages that the online marketplace can provide (Laudon and Traver, 2007). Basically, e-commerce is commerce enabled by Internet technologies, including pre-sale and post-sale activities (Whiteley, 2000; Chaffey, 2004). Since the late 1990s, e-commerce's rapid growth is obvious in the developed world. Global e-commerce spending has currently reached US$10 trillion and was US$0.27 trillion in 2000 (Kamaruzaman et al., 2010). The United States, followed by Europe, constitutes the largest share with about 79% of the global e-commerce revenue (ibid.). However, the African and Middle East regions have the smallest share with about 3% of the global e-commerce revenue (ibid.).

In Saudi Arabia, e-commerce is still a new wave in the country's information technology revolution. Although Saudi Arabia has the largest and fastest growth of ICT marketplaces in the Middle East (Saudi Ministry of Commerce, 2001; Alotaibi and Alzahrani, 2003; U.S. Commercial Services, 2008; Alfuraih, 2008), e-commerce growth is not progressing at the same speed (Al-Otaibi and Al-Zahrani, 2003; Albadr, 2003; Aladwani, 2003; CITC, 2007; Agamdi, 2008). Only 9% of Saudi commercial organizations, mostly medium and large companies from the manufacturing sector, are involved in e-commerce mainly Business to Business (B2B) (CITC, 2007). Several studies have been conducted to discover the reasons behind the slow e-commerce developments in the Arab world in general and Saudi Arabia specifically. The reasons mainly cited are poor ICT infrastructure, trust and privacy issues, cultural issues, and the absence of clear regulations, legislation, rules and procedures on how to protect the rights of all involved parties (Albadr, 2003; Aladwani, 2003; Al-Solbi and Mayhew, 2005; CITC, 2006; Alfuraih, 2008; Agamdi, 2008; Alrawi and Sabry, 2009; Alghaith et al., 2010).

The remainder of this paper is laid out as follows. Section 2 reviews the extant literature regarding online retailing in other countries, retail business in Saudi Arabia, and e-commerce in Saudi Arabia. In section 3, our research methodology is explained. Sections 4 and 5 present what Saudi retailers see as inhibitors and enablers to online retailing in Saudi Arabia. In section 6, we present an analysis of the data, the limitations of the research so far and directions for future work. We conclude in section 7 with the research implications and conclusion.

## 2.   LITERATURE REVIEW

In this section, we review the extant literature that defines and explains the e-commerce phenomenon. Literature is reviewed regarding global online retailing and domestic e-commerce in Saudi Arabia. Finally, literature focusing on the nature of retailing business in KSA was reviewed to highlight the unique nature of Saudi retail context.

### 2.1.   Global Online Retail and E-commerce background in Saudi Arabia

Basically, online retailing is an Internet enabled version of a traditional retail system. It includes four sub-types: Virtual Merchants (online retail store only); Bricks-and-Clicks e-retailers (online distribution channel for a company that also has physical store); Catalog Merchants (online version of direct mail catalog); and, Manufacturers selling directly over the web (Laudon and Traver, 2007). It is classified as Business to Customer (B2C) e-commerce. B2C represents a small share of e-commerce revenue compared to Business to Business (B2B) e-commerce throughout the developed world (Kraemer et al., 2006). The USA followed by the UK account for the world largest market for online retailing. Online retail in USA accounts for 3.6% ($142 billion) of total retail sales in 2008 (U.S. Census Bureau, 2010) and in UK accounts for 10.7% (almost $74 billion/52 billion Eur) of UK retail trade in 2010 (Centre for Retail Research, 2010). According to the Nielson (2010) report, the top 10 products/services globally sold online are books, clothing/accessories/shoes, airline ticket/reservations, electronic equipment, tours/hotel reservations, cosmetics/nutrition supplies, event tickets, computer hardware, videos/DVDs/games, and groceries.

A study was conducted as a cross-country comparison; covering USA, Brazil, Mexico, Germany, France, Denmark, China, Taiwan, Singapore and Japan, to "examine the key global, environmental and policy factors that act as determinants of e-commerce diffusion" (Gibbs et al., 2003). It finds that B2B e-commerce seems to be driven by global forces, whereas B2C seems to be more local phenomena. The determinants which act as drivers/enablers for B2C e-commerce are: consumer demand to buy online; business desire to expand a market or compete; consumer purchasing power; strong ICT infrastructure; and, government promotion. The factors which act as barriers/inhibitors for B2C e-commerce include: lack of valuable and useful content for consumers; inequality in socioeconomic levels; consumer reluctance to buy online; lack of trust due to security/privacy concerns; consumer preferences for in-store shopping; existence of viable alternatives, such as dense retail networks, convenience stores; lack of online payment options, lack of customer services; and, language differences (Gibbs et al., 2003). Ho et al. (2007) proposed three related theories to describe the underlying mechanism for growth in e-commerce revenues at the national level. They studied 17 Western European countries over five years (2000-2004). Their study reveals factors that have significant impact on a country's B2C e-commerce growth on two different market levels: endogenous and exogenous. On the endogenous level, the key factors are: Internet user penetration; telecommunications investment intensity; and, education levels within a country. On exogenous market level the factors include: the extent of Internet user penetration in a country; the intensity of telecommunications investment; the availability of venture capital; general education level; and, the degree of credit card penetration in the economy.

Kendall et al. (2001) explored factors that influenced Singaporean small and medium enterprises (SMEs) to adopt e-commerce. They used Diffusion of Innovation (DOI) theory as a framework; however their usage of this theory was incomplete because they focused only on technology attributes and ignored the other contributing factors that affect the diffusion of a new technology. In any case, the relative advantage of using e-commerce in their businesses emerged as the most significant factor influencing their willingness to adopt such technology. Forrester Research Inc (2010) surveyed 114 Australian online retail professionals. The results show that there are common challenges; however, they affect some type of firms more than others. Larger firms and multichannel retailers face greater challenges with finding and maintaining a force of skilled employees. Smaller firms face greater challenges with logistics, and individuals face challenges with marketing. Access Economics Pty Limited (2010) reported: lack of understanding of online business models; set-up and ongoing maintenance costs; limited and/or unreliable broadband access in some areas; complexity; and, currency and payment issues, as all concerning domestic retailers in Australia in the adoption of an online channel. Paying tax for online purchases in Australia may represent a barrier for some retailers. Some retailers such as Myers and Harvey Norman have set up their own online, offshore stores in China in order to avoid charging their customers with GST in Australia (Sexton, 2010, Speedy, 2011). In contrast, the tax-free environment for online sales/purchases encouraged the online retailing growth in USA (Dinlersoz and Hernández-Murillo, 2005). Wymer and Regan (2005) identified 26 factors from the literature that may influence a decision to adopt e-commerce by SMEs. They then consolidated these factors to determine their level of influence, either positive or negative. A survey used for this study and geographical location for this study was in the Appalachian counties of Kentucky in USA. The study revealed 16 factors were found to be significant, ten as facilitators and six as impediments. Incentives included: need; innovativeness; competitive pressure; value; government promotion; reliability; available e-commerce technology; effective business models; prior experience; and, executive experience. On the other hand: set-up and maintenance cost; priority; security; available capital; market; and, partner/vendors were considered as barriers.

In Saudi Arabia, so far, the effort towards e-commerce development have not reached its originally stated aspirations; neither what it sees as the world's expectations of a country of the level of importance and weight in the global economy like Saudi Arabia. Official/government information on e-commerce in Saudi Arabia is poor. Since 2006, the responsibility of e-commerce has transferred from the Ministry of Commerce to the Ministry of Communications and Information Technology. This information was gained from an exploratory phone call made in December 2010 to the Saudi Ministry of Commerce. It was explained that the Ministry of ICT in Saudi Arabia is still in its early stages of studying e-commerce. Currently, they are conducting a survey on e-commerce in Saudi Arabia and a report may be published in May/June 2011.

Firms in Saudi Arabia seem not to be following the developed countries' rapid progress towards global e-commerce. In contrast, online shoppers in Saudi Arabia are increasing with access to technology and communications infrastructure. The Arab Advisory Group carried out an extensive survey in mid-2006, targeting Internet users in four Arab countries (Saudi Arabia, UAE, Kuwait and Lebanon). The survey covered Internet usage and, e-commerce activities in these countries. While UAE ranked first in the rate of annual spending on e-commerce per capita, Saudi Arabia ranked first in the overall money spent on e-commerce activities. As for the prevalence of e-commerce activities among the population, UAE ranked first at 25.1%, Saudi Arabia second at 14.3%, Kuwait third at 10.7% and Lebanon last at 1.6% (AAG 2008). A recent survey of Saudi Arabia's 11.4 million Internet users (representing 41% of population) found that around 3.1 million Saudis have purchased online. Airline tickets and hotels bookings take the largest percentage of these purchases (ACG, 2009, AAG, 2011).

Some studies have been conducted to investigate the challenges of e-commerce in Saudi Arabia. These challenges involve the absence of clear e-commerce regulations, legislation, and rules (Al-Solbi and Mayhew, 2005; Agamdi, 2008). Although Saudi Arabia contributes to the efforts of UNCITRAL (United Nations Commission into International Trade Laws) (Saudi Ministry of Commerce 2001), there is a need to have major development in terms of e-commerce regulations, legislations and rules to protect the rights of all parties involved in e-commerce transactions (Albadr, 2003; Al-Solbi and Mayhew, 2005; Agamdi, 2008). Other challenges involve the domestic mailing system (Alfuraih, 2008). Before Saudi Post was established in 2005, individuals had no home addresses (Saudi Post, 2008); therefore, to receive mail, individuals had to subscribe to have a mailbox in the post office (Alfuraih, 2008).

## 2.2.    Retail Business in Saudi Arabia

The average of annual population growth rate in the Kingdom of Saudi Arabia has reached 3% over the past 10 years. This seems to be playing a role in making Saudi Arabia the most dynamic retail markets and most notably in the Middle East (AMEinfo, 2008). "With an estimated population of 24.9 million and a per capita GDP of US$24,581 in 2008, Saudi Arabia is the largest and one of the richest retail markets in the Middle East" (ACG, 2009). This is illustrated below, in Figure 1.

According to the latest report issued by Saudi Alhokair group, the wholesale and retail trade grew at a compound annual growth rate of 5.8% in the past 10 years. In 2010 the retail trade volume exceeded SAR90 billion (US$1= KSR3.75), although it was only expected to reach up to 70 billion. It is expected that the volume of retail trade to SAR 130 billion in 2012 (AMEinfo, 2008). In 2011, the size of the retail market in Saudi Arabia estimated greater than KSR160 billion dominated by small and medium size companies accounting for more than 85% of market share (Habtoor, 2011).

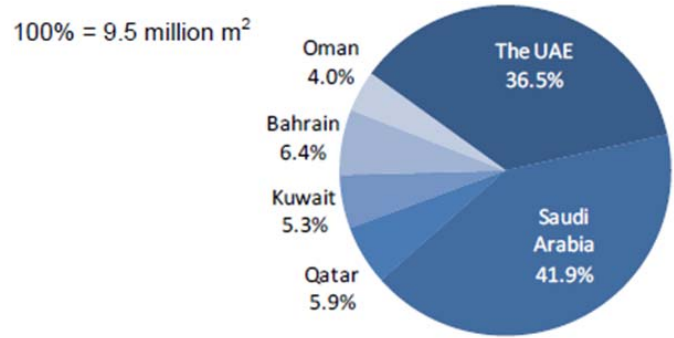

Figure 1: Completed Gross Leasable Area in the Gulf
Countries, 2008 (%), Adapted from (ACG, 2009, p 22)

According to the Saudi Ministry of Municipal and Rural Affairs 2010 statistics, the number of retailers in Saudi Arabia reached 242 thousand; 54 thousand of them are licensed for grocery (Habtoor, 2011). The retail sector in the Saudi market is fragmented and dominated by sales in individual stores, despite the emergence of a number of large retail chains in recent years (AMEinfo, 2008). "Most of the retail industry in Saudi Arabia is dominated by unorganized retailers like Baqalas (neighbourhood stores) with the top five retailers accounting for only 13.9% of the total market. The opportunity for the organized retailers to increase their market share is therefore vast" (ACG, 2009). See Figure 2.

| No of Outlets (% industry turnover) | Saudi Arabia | The UAE |
|---|---|---|
| Hyper/supermarket | 430 (39%) | 280 (85%) |
| Other self-service | 3490 (20%) | 260 (3%) |
| Large grocery | (10860) 22% | 1550 (6%) |
| Baqalas | 18150 (18%) | 5040 (7%) |

Figure 2: Trade Structure
Adapted from (ACG, 2009, p 23)

Baqalas (neighbourhood stores) are almost in each main street in Saudi main cities. These types of shops are located at the same residential/business buildings. A typical retail street scene is depicted in Figure 3.

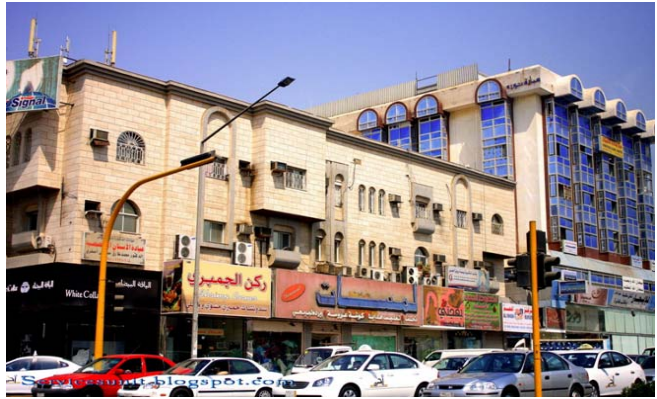

Figure 3: Baqalas (neighbourhood stores) along Jeddah
Road
Source: (BWN 2008)

## 2.3.   Significance of this Study

While the literature provides an extensive list of issues that influence the adoption and use of electronic commerce by commercial organizations in the western and Asian countries, the literature about these issues in the Middle East and Africa is poor. There is a lack of published empirical studies of e-commerce relating to the Arab world. Similarly, a qualitative study surveying retailers in Saudi Arabia does not exist. Much of the research up to now has been descriptive of e-commerce in Saudi Arabia and has not gone into deep analysis regarding the adoption and diffusion of e-commerce. In addition, no research has been found that focused on one model of e-commerce to investigate rather than e-commerce in general. This study starts to address each of these issues.

The topic e-commerce is broad and involves several different and distinct models (e.g. B2B, B2C, C2C, B2G etc) and as such makes it difficult for a single comprehensive study to be undertaken. Investigation in one area of the topic, in this case B2C, helps to clearly define and emphasize the effects on all involved parties and avoids the dispersal of the subject that occurs when it is broadly addressed. For this reason, this study focuses on online retailing as a model of Business to Customer (B2C) e-commerce system and fills a gap in the extant literature. Updating primary data about retailers' concerns and suggestions regarding online retailing in KSA is another important outcome of this study. Published studies of e-commerce in Saudi Arabia are now at least three years old. Considering the speed of diffusion of ICTs, the Internet, and e-government related applications in the KSA, there is a significant need maintain currency with existing issues. The Internet users' boom in Saudi Arabia has reached to 41% of the population by the end of the third quarter of 2010; where it was only 13% in 2005. Similarly, broadband subscriptions have grown from 0.06% of the population in 2005 to over 12.2% at the end of the third quarter of 2010 (MCIT, 2010). Also many existing studies predate the new free service offered by Saudi Post for home mailing. The final significance of this research emerges from highlighting the missing role of Saudi government as a promoter of e-commerce development. While government promotion has limited effects on the diffusion of e-commerce in most countries (Kraemer et al., 2006), this study significantly indicates government promotion and support as a key driver to online retailing in KSA.

## 3.       RESEARCH METHODOLOGY

This study initially involves exploratory research using a qualitative approach. By adopting a qualitative approach this study is able to gain an in-depth understanding of the e-retail phenomena in KSA. A semi-structure interview was designed based on the variables determining the rate of adoption of innovation (Rogers, 2003). Rogers (2003) identified five attributes determining the innovation's rate of adoption. He highly recommended that each diffusion research should develop the measures of the five perceived attributes. These five variables are (1) perceived attributes of innovations (Relative advantage, Compatibility, Complexity, Trialability, and Observability), (2) type of innovation-decision (optional, collective, authority), (3) communication channel diffusing the innovation at various states in the innovation-decision process (mass media, interpersonal), (4) nature of the social system (norms, degree of network interconnectedness, etc), and (5) extent of change agent's promotion efforts (Rogers, 2003, p 222). Based on these five attributes, interview questions were designed. The interview questions are provided in Appendix. Furthermore, a free time was given at the end of each interview session to allow the interviewees - if they wish - to add further information or comment.

**Table1: The Sample Demographic Profile**

| Business Category | Participant's Position | Company Size | Have access to Internet | Have a website? | Sell online? |
|---|---|---|---|---|---|
| Appliances and home improvement | CEO | Large | ✓ | ✓ | ✗ |
| Beauty products | Marketing Manager | Large | ✓ | ✓ | ✗ |
| Internet Service Provider | Director | Large | ✓ | ✓ | ✓ |
| Sporting Goods | Regional Director | Large | ✓ | ✓ | ✓ |
| Telecommunications | Regional Director | Large | ✓ | ✓ | ✗ |
| Books | CEO | Medium | ✓ | ✓ | ✗ |
| Chocolates | IT Manger | Medium | ✓ | ✓ | ✓ |
| Groceries | Director | Medium | ✓ | ✓ | ✗ |
| Printing Services | Director | Medium | ✓ | ✗ | ✗ |
| Watches | Owner | Medium | ✓ | ✓ | ✗ |
| Appliances and home improvement | Owner | Small | ✗ | ✗ | ✗ |
| Audio/Video | Owner | Small | ✗ | ✗ | ✗ |
| Computer related | Director | Small | ✗ | ✗ | ✗ |
| Groceries | Owner | Small | ✗ | ✗ | ✗ |
| Kitchen utensils | Director | Small | ✗ | ✗ | ✗ |
| Mobile Phones | Owner | Small | ✓ | ✗ | ✗ |

The researcher officially sent, in person and mail, official letters to more than 40 retailers in Jeddah in Saudi Arabia. However, this plan was failed where no responses were received. The researcher then tried to contact the relevant departments (in government and business sectors) to provide help in order to contact retailers. They were not helpful in this regard. They refer the researchers to their websites to find out retailers' contacts. For this reason, the researchers resorted to special relations to coordinate the selection of participants for interviews. The plan was to conduct interviews with 20-30 retailers, however as result of facing this difficulty the number of interviewees is 16. The sample covered various categories of retail market. Interviews were conducted with 16 retailers' decision makers (owner, headquarter manager, marketing manager, and IT manager) in Saudi Arabia. The sample were selected to cover large, medium and small companies and also to cover different types of retail businesses (telecommunications, computers, sports, super markets, restaurants, printing services, Internet services, electrical and electronic products, beauty and body cares, books, watches and clocks, and Chocolate and Biscuit manufacturing). See also Table 1. Each interview session took a time up to 30 minutes. In this paper each participant is given a name. The names mentioned in this paper are not real and are given for identification purposes while keeping participants' anonymity.

## 4.      CHALLENGES FACING RETAILERS TO ADOPT ONLINE RETAILING CHANNEL

In this section, we present the challenges encountering retailers in Saudi Arabia to adopt online retail channel. Content analysis was used and 12 factors which believed to be challenged are identified. Under each identified factor, direct quotes from the participants' speech are used as proof of that the identified issue is a challenge. The discussed challenges are culture of people not supporting for online sales; difficulty to offer competitive advantage on online channel; difficulty to gain profit from online channel; cost of setup; type of products are not suitable to be sold online; lack of clear e-commerce legislations in the country; lack of e-commerce experience; lack of online payment options help to build trust; delivery issues; trust and security issues; resistance to change; and poor ICT infrastructure.

### 4.1.    Culture of People not Supporting Online Sales

Basically culture can be defined as integrated set of knowledge, beliefs, values, attitudes, goals, behaviors, and practices that characterizes a group or society of people and determines their way of life (Morsi, 1995). The culture of people in Saudi Arabia is the most influencing factor that affects retailers' decisions to adopt e-retail systems. This is the first answer that you get when you ask a retailer in Saudi Arabia why you do not apply e-commerce system in your company. There is an emphasis on this factor making it a key concept for discouraging the growth of e-retailing systems in KSA. Mr. Mohammed commented "You see in this part of the world we haven't had a strong e-commerce movement. I originally came from the western part of the world and their day to day activities: to go and check online and buy things online, it provides comfort, convenience, and gives the ability to check prices and compare them as well. But unfortunately we have not seen a strong foothold of e-commerce in this part of the world". Mr. Ahmed said "the percentage of users who browse the Internet is high, however those who sell or buy is very small. Culture of the people to accept e-shopping is not encouraging us to open the electronic market... Recently we have bought powerful software, but unfortunately the culture of our customers is not encouraging us to go ahead with this idea. Customers do not trust to buy online and they only accept with strict conditions, because it requires credit cards payment. Recently the idea of having Internet payment cards has increased. There is International website where you can buy from and delivered to your home by International delivery companies like DHL, but internally in Saudi Arabia so far is not encouraging". The culture of people in this country is to go to shop and "see the products physically and do inspections by themselves before buying", Mr. Saleh said. "People here are not familiar with e-commerce and this is not something that you can apply in a day. People here have to come to the shop to see the product and inspect it", said Mr. Ali. Mr. Nasir said: "Selling online like Amazon.com is not useful for us in this country and all Arabic countries. This subject refers to peoples' culture of selling and buying. So the normal way of business is the best for us".

Mr. Nasir said: "Honestly there is ignorance inside our company in terms of e-commerce. Also there is ignorance in the community. Most of them don't know the meaning of e-commerce as result they will not venture. There are people when they pay online remain in doubt and not sure of receiving their purchased orders. There is fear, because ignorance breeds fear. There is doubt about dealing with money on the Internet". Mr. Ali said: "The majority of people in our country do not know how to buy and sell on the Internet. The ignorance of something makes it difficult". However, Mr. Mohammed totally disagrees to blame customers that they are not educated in this field. He said: "People here understand the benefit of e-commerce 100%, when I was in Riyadh I visited GYTEX (a technology fair) and the people I saw there, the indigenes, original Saudi people were superbly well educated and they knew exactly the benefit of e-commerce and everything. But I think someone needs to kick off the whole thing of having the access of Wi-Fi, of having the access of broadband, having the access in every house hold. Once this done, obviously you will see the results of that very soon". Mr. Osam supports that Saudis are not ignorant in this field and he places significance on the fact that e-commerce is fairly new in the Saudi environment. Mr. Waleed said exactly the same thing: "We are growing community, and the idea of e-commerce is new to us, and we need time to be more familiar with these things".

The habit of buying online from e-retailers in Saudi Arabia is not familiar and this is confirmed by three participants whose companies have e-retailing systems. All of them have confirmed that the performance of online sales cannot be compared in any way to the normal way of selling. Mr. Salem said: "This region doesn't give you more sales on the Internet because the habit of the people here in this region is to go outside and buy, we are not in Europe or America where people easily buy from the Internet, culturally we are different than them that's why e-commerce is not getting much to more businesses". Even with offering discounts and free offers for those who buy online there is no demand. Mr. Thamer's company offers "10% off on online sales" and they provide free delivery service inside the main cities however "there is no good demand to buy online. For example, in western region of Saudi Arabia we receive 3-5 orders buying online every week and these are nothing compared to normal way of business. But we will continue providing this option (selling online) to encourage people to use it". Mr. Thamer mentioned that: "there is specific class in the community, according to a study we conducted 6 months ago, that prefers to buy online. This class has a previous experience studying overseas in western countries and gets used to buying online. They have become comfortable with online purchases".

## 4.2.    Difficulty to Offer a Competitive Advantage on Online Channel

Having a competitive advantage is very important to be successful gaining profit share in the market. Some participants in this study showed their concerns that they cannot offer advantage for their customers online and as result they will not gain more profit. The participants who have concerns regarding this issue mainly are because the delivery fees if there is any. "If you request from the customer to pay extra 10 SR for such a fees like this and if the price in total comes more than the price in the local shops, he/she may buy it from them without buy it from you online". A marketing manager of a company selling perfumes and beauty products said: "I can see also one of the inhibitors is delivery fees or insurance. If the products available in the local market with the same price then the company which offer selling these products have no competitive advantage". An owner of a computer shop said: "There might be extra fees to buy online while computer shops are widespread in Jeddah, so to go ahead with this idea is not benefiting us". An owner of electrical and electronic goods company said: "I agree to go ahead with this idea if there are products that are not available in the local market where we can have competitive advantage. However electronic and electrical shops are almost in each street in Jeddah which makes the idea of e-commerce not useful for us".

This is confirmed by the owner of a manufacturing company producing Islamic watches and clocks. The company has e-commerce website mainly targeted for international market. This is because, as the owner said, "Islamic watches and clocks that we produce and sell are unique products in the international market. The purchasing power comes from overseas and that's why this website is in English. We don't target consumers in Saudi Arabia and we don't have a priority to do so. This is because all our products are plentiful in the Saudi marketplace. So, we don't have a competitive advantage in Saudi market to encourage locals to buy online".

This is an interesting and important point for discussion. The usual way around this is to think of different "online" business models with efficiencies to reduce the cost of doing business. It is much like KSA moving from petrol to information economy as retailers must reduce the reliance on transport (cost of petrol and time) and storage of stock (rental). For example, Amazon can sell books cheaper than many local bookstores because it does not have to rent shop space or run a delivery service itself. Order details are sent directly to publishing houses for direct delivery by a specialist courier or postal service. There are also models that provide certain items only by Internet sales so customers have to use that channel if they want that product. Similarly, access to specialist items direct from manufacturers or wholesalers is a price reduction model. Part of the problem may be that Saudi retailers are not aware of the need to adopt different business models to take advantage of different forms of disintermediation – removing service costs from the retail price. Many new companies in the West have developed and implemented delivery businesses specifically for difficulties around online orders. This is part of the "education" problem surrounding e-commerce – not just customers but for the retailers. Saudis are not adept at creating the competitive advantage online yet. It is a new way of thinking about supply chains.

## 4.3.    Delivery Issues

There are some companies in Saudi Arabia having their own delivery system. Also there are some others have no problem with the delivery for orders by phone. They organize with local shipment companies to deliver the goods, however, this is not considered as a professional delivery service because these normal delivery companies do not get a clear home address and that's why they deliver to their offices only and ring the customer to come to pick up his/her order. When a customer wants the products to be delivered to their home they have to arrange with the delivery company and pay an extra fee. They have to explain to the driver where their home is located. Mr. Waleed's company offers orders by phone and the money should be deposited in their bank account in order to be delivered. They then arrange with a local shipment company called Alzagel to deliver the orders "with 10 SR delivery fees". They "only take the name and mobile phone number of the customer. When the order arrives in the shipment office in the customer's city, the shipment company calls them to come to take their product from the office but if they cannot come then they require payment of extra fees to deliver to their home". Mobile phone contact is very important in this case because the driver and customer have to be in contact till they have received the delivered products. Mr. Salem said: "We depend on the mobile phone number; if the mobile phone number is not correct then it's difficult to deliver the product. This is not only for us but also for other companies and home delivery people they are struggling with these things. But there are certain locations (e.g. companies' buildings) you can know the address clearly, you know which street and block, building, floor, flat etc. But with community houses is not very clear. So we are managing well with the mobile phone numbers actually. By calling the customer to describe to us their location we will deliver the products to them".

However, there are other issues involved with the delivery. Some participants are concerned that their products are sensitive and need high care in the delivery which means more money to ensure that the products are delivered in satisfying condition for the customer. Mr. Tameem asked: "who will pay for guarantee fees the seller or customer? If you request from the customer to pay extra 10 SR for such a fee like this, and if the price in total comes more than the price in the local shops, they may buy it from them rather than buying it from you online". Mr. Ahmed and Mr. Hassan have the same question and answered similarly: "customer maybe will not accept to pay delivery fees".

Some of these issues were encountered elsewhere and were eliminated through economies of scale. Insurance companies can offer large discounts to companies that bring them a lot of relatively risk free business. This reduces guarantee costs. Some of the costs for doing online business have been covered by the credit card companies for companies reducing transaction errors and delivery practices that eliminate non-repudiation. Participation in "online shopping malls" allows several businesses to share the services of one company in order to get bulk discounts for things like delivery and insurance. Again it is about business models that suit the online marketplace and the local conditions. To be honest, businesses are not thinking beyond their traditional models if they cannot find ways to save enough in their Cost of Goods Sold to cover delivery cost.

## 4.4.    Having Few Online Payment Options that Help to Build Trust

The main issue involved with online payment is trust. How can a customer trust an e-retailer when they are paying online? Mr. Salem said: "We notice that people feel it is difficult to put their credit card information online. It's okay if it is Amazon.com. There is no problem because Amazon for them is very trustworthy name but for us they will feel it is very difficult to buy! There is a threatened feeling wondering if they will get the products or maybe not get the products". Mr. Moneer agrees with Salem that: "people in general tend not to use credit cards to buy online except for those that are used to doing it. However people here are afraid to buy online".

Mr. Nasir's company offers information about books and their prices on their website. It is possible to do online orders but there is no option for payment. When doing an online order a customer has only one option which is to leave his contact details and a message appears stating that they will be contacted soon. However, Mr. Nasir said: "We do not use the website to sell online because the online payment is not available. We only have a bank account number that we give to those who would like us to deliver their orders to home. We receive their orders, by fax or e-mail, and we give them our bank account number to deposit the money. As soon as we receive the money we send the books with delivering companies like FedEx, DHL, or other shipment companies". When they asked why they were not offering online payment they replied "we don't trust online payment and our customers don't trust it either. The ignorance of something breeds fear of it. There are thefts of credit cards numbers, and there are hackers that penetrate your privacy. So this is a problem for the money dealing on the Internet".

Much of this was also experienced in the West and the banks started making "debit cards" available that could have a low set limit prepaid by cash transfer to the card account. This reduced the possible exposure to theft and fraud allowed with credit cards. As the payment systems mature and as each stakeholder: buyer, seller, bank, and credit card company, work out how to limit risk to an acceptable level for them, so the online payment system gains a hold. I believe that trust networks, certification authorities (CAs) and public key infrastructure (PKI) to secure transactions all play a huge part in online business success. Mr. Mohammed suggested that there should be a mechanism for online payment: "whereby the customers feel confident" and help to build their trust with an e-retailer. Mr. Naif, whose company has an e-retailing system that targets the international market, strongly agrees with Mr. Mohammed that having a mechanism for online payment to build the trust with the customers is helpful. He sees that: "not affording more options for online payments" affects the decision to adopt e-retailing system. We need to have more options for online payment to help customers to trust the e-retailing system and to encourage them to buy online, "people are afraid to buy online using credit cards", Mr. Osam said.

## 4.5.    Poor ICT Infrastructure

Infrastructure that helps to move towards e-commerce is not forthcoming in Saudi Arabia. "Unfortunately we have not seen a strong foothold for e-commerce in this part of the world. But, I'm confident because the market is so buoyant out here... So I have a strong conviction that people will actually go into e-commerce, but we need to have the infrastructure right. In order to have the infrastructure right we must make sure that Wi-Fi is available in better cases and more places. We must ensure that broadband is available for other people in remote places and we must also ensure that the protected payment plan is also taken care of, so that we do not create a system where people actually pay and do not know whether or not the goods will arrive. So it is very important to get the infrastructure right". Mr. Mohammed said.

Mr. Thamer said: "To use the Internet for selling and buying we must have infrastructure right. There is a movement and growth toward e-commerce but it is very slow". Getting infrastructure right is very important and must be ensured "to correctly apply e-shops". When we talk about infrastructure we mean: "(1) The telecommunication networks inside the city and the country to be at high speed and high efficiency; (2) This must be supplied, as much as possible with minimal interruption (high level accessibility). If a customer made an online purchase and they don't receive the order in the expected time they will not be satisfied and trust will be affected. The problem may be not with the delivery system, it may be with the communication system itself not sending or making a delay in the order request. (3) Having an integrated system such as the Saudi Stock Market. If there is something similar for e-commerce it will be great. Having an integrated system for e-commerce is very important in terms of having people take care of the system, search for problems and solutions, and leading development. I believe that the Saudi market is bullish and a fertile ground for investments", however this needs time and requires a lot of effort "to reach the maturity in this field".

## 4.6.    Lack of E-commerce Legislation in KSA

Regulations and rules are very important in terms of systemizing the work and protecting rights for all involved parties in commercial transactions. When there are no rules or the law for specific kinds of business is not clear of course there will be a reluctance to move forward. Mr. Ahmed commented regarding having clear law for e-commerce: "there is no system... It has to be done from an early stage. Customers have to be ensured that there are rules and legislations to protect their rights. This is very important to build the trust with customers. If a customer pays and no products delivered this means the seller is branded a liar! It has to have a clear system, i.e. from government, imposed on all involved parties". Mr. Waleed agreed and he commented: "We are a society subjected to several issues of fraud, such as stock market radical decline, Sawa cards, lands and fake projects! We did not find a clear law to protect our rights in these issues. Now if we go ahead with selling and buying online the same problem happens in the absence of clear laws and regulations. With no clear e-commerce law and legislations applying to e-commerce systems it is not encouraging. In the future the environment may be encouraging, but currently the situation is not".

This is part of a problem called repudiation of participation in a transaction. Yes, if a customer pays and no items are delivered then there are questions asked about the reliability of the seller or the delivery system. On the other hand, if the goods are delivered to an address but the customer says that they did not receive them and does not pay, then it is the customer or the delivery process that cannot be trusted from the sellers' perspective. Payment systems need to have parties to the transaction identified so that funds cannot be withheld or so that goods are not delivered. This is the realm again of CAs and PKI. An effective delivery tracking system also helps with this process to keep track of parcels.

The oldest commercial rule of all is: "Caveat emptor" (Latin for "Buyer beware!") but different countries have different laws regarding things like marketing and advertising that limit scams that can be levelled against participants in transactions.

## 4.7.   Lack of Trust

Some participants do not trust online systems to sell goods. This concern is involved with online payment. Mr. Nasir clearly said: "We don't trust online payment and our customers don't trust it too. The ignorance of something breeds fear of it. There are thefts of Credit Cards numbers, and there are hackers that penetrate your privacy. So this is a problem for the money dealing on the Internet". Mr. Ali has the same concern as he: "does not trust online payment, the money must paid cash". Mr. Waleed commented: "the Internet is fertile ground for cheating and fraud. I think the acceptance to sell online will remain weak".

## 4.8.   Type of Business/Product are not Suitable to be Sold Online

Type of business or product plays a role for some companies to take the decision to adopt an e-retailing system or not. For example, companies which sell food and fragile products feel that selling online is not good due to their concerns regarding delivery the order in a quality condition. In a city where the temperature reaches 50-55 degrees in the summer, delivering food: "needs special care because it requires being stored at a specific temperature to be delivered safely and healthy". Fragile products are sensitive too because customers are afraid that the product maybe will not be delivered safely. A manufacturing company of chocolates and biscuits has conducted a small research in this area in the late of 2009. They found that: "the fragile products are sometimes losing the deals! Some customers are interested to buy from you but when they see fragile products they will go away. It is better for them to buy from local market instead of buying online where they cannot ensure the product will be shipped safely".

There are other issues involved with the type of product other than issues involved with the delivery. For example a beauty company selling perfumes, makeup, cosmetics and body lotions, shampoos and skin care: "have tried to market on the Internet but there was no demand because our products related to smell, and shape, and the customer wants to see the physical product in front of them". Also the nature of the business may affect the decision to adopt online selling. For example, a company selling electrical goods (e.g. fridges, air conditioners, TVs, ovens etc.) and electronics equipment and tools finds it very difficult to even think of selling online because "specifically the nature of our business makes it difficult to go ahead with this idea. Some products are widespread in the market. Yes I agree to go ahead with this idea if there is products are not available in the local market where we can have competitive advantage. However ... the majority of similar companies, 99%, do not have e-commerce websites ... electronic and electrical shops are almost in each street in Jeddah which makes the idea of e-commerce not useful for us".

## 4.9.   Lack of E-commerce Experience

Certainly ignorance is something that generates fear of use. When Mr. Tameem was asked what his company needs to do in order to adopt the use of an e-retailing system, he replied "I don't know the procedures". In another context when he asked about the difficulty of using e-retailing systems, he replied "I feel it is difficult and causes a headache. The normal way is better and easier". Mr. Nasir clearly said: "Honestly there is ignorance inside our company in terms of e-commerce. Also there is ignorance in the community". And Mr. Ali said the same thing: "The majority of people in our country do not know how to buy and sell on the Internet. The ignorance of something makes it difficult… we do not think about e-commerce at all. The normal way of selling is the best for us and we do not want to sell online at all".

### 4.10.    Difficulty to Gain Profit through Online Channels
There is no doubt that the main target for businessmen is to gain more profits. When there is an idea that seems not helpful in terms of gaining more profits, it has more chance to be rejected. Mr. Saleh stated: "Our target, as businessmen, is to gain more profit but when there is an idea that is not helping us to gain this goal then there is no need to apply or think about it". When Mr. Hassan was asked if there are difficulties in adopting an e-retailing system, he replied: "For us there is no difficulty but a manufacturing company (like Alsaef) which we import products from, they have their own website. Take this example, Alsaef's products already demonstrated on the manufacturing company website, so there is no benefit for us to reinvent the wheel!" "This region doesn't give you more sales on the Internet because that is the habit of the people", Mr. Salem said.

Mr. Nasir said: "Although there are more things that you can save when you transfer to selling online like tools, equipment, and rent of shops, I don't think that the Internet will help us to achieve more profit... Selling online like Amazon.com is not useful for us in this country and all Arabic countries. This subject refers to peoples' culture of selling and buying... The normal way of selling remains the best for us to gain more profit".

### 4.11.    Resistance to Change
It's normal to find resistance for a new technology or idea as this needs time to become familiar with and then to be accepted. Rogers (2003) shows an adoption timeline for a new technology/idea where few adopters come in the beginning and with time the percentage of people adopting use increases. Some participants in this study don't want to change because they find the normal way of selling is much better and more profitable for them. Mr. Saleh said: "The normal way of our business is better. People are used to seeing the products physically and do their own inspections before buying. Currently this is not benefiting us". Despite this Mr. Fadi's company has orders by phone and its own delivery system. They reject the use of an e-retailing system at the moment as they are happy with: "Using the Internet to order our products is not our priority at the moment and we have not thought about it or done a study to find out what needs to be done". Mr. Ali said: "This depends on familiarity. We are familiar with the way of normal selling where a customer has to come to our shop and see the products... It is much better for us when a customer comes to the shop... We prefer personal marketing it is much better for us. When you explain to a customer face to face is totally different from leaving him alone to read the features of a product on the Internet... As result, we do not think about e-commerce at all. The normal way of selling is the best for us and we do not want to sell online at all... in our business of selling electronic and electrical products the normal way is the best".

### 4.12.    Setup Costs
For some participants, the setting up cost to build e-retailing system represents a difficulty. They see that there are a lot of requirements, e.g. training programs, computer network etc. that need to be done prior to applying the system. When Mr. Saeed was asked if they find it confronting to use online systems, he replied "yes, you have to design a commercial website, train your employees, deliver goods". Mr. Waleed has concern with setting up cost and is afraid to spend money for something that is not beneficial. "Financial costs are involved in something that does not bring profit. As I told you earlier this type of business is not useful currently and returns very low profits".

This is one of the aspects that taxation law is useful for in many countries. If e-Commerce is a priority then the country can create tax concessions to make it less of a burden and to cover the initial losses that may be incurred. For somewhere like the KSA where there is no taxation then concessions can still be paid to offset business losses during adoption. This is where government support is important to promote the adoption of technology.

## 5.    ENABLERS FOR ONLINE RETAIL GROWTH IN SAUDI ARABIA

Developing e-retailing systems growth requires a lot of effort to create an enabling environment in a country where selling and buying online seems to be an innovation. Rogers (2003, p 12) define the term 'innovation' as "an idea, practice, or object that is perceived as new by an individual or other unit of adoption". From the previous section it is apparent that retailers in Saudi Arabia are disinclined to adopt e-retailing systems at the moment. The inhibitors that were identified appear to provide a commercial environment that is not currently conducive to e-commerce. By providing solutions that address inhibitory factors it will be possible to create a positive online business climate. Through this exploratory research there also arose several enabling factors that would promote e-retailing systems growth identified by retailers in Saudi Arabia. These include educational programs and building the awareness of e-commerce, government support and assistance for e-commerce, trustworthy and secure online payment options, develop strong ICT infrastructure, and provision of sample e-commerce software to trial.

### 5.1.    Educational Programs and Building Awareness of E-commerce

Education and building awareness of e-commerce has received high attention by the participants. They consider them as key factors toward the diffusion of e-retailing businesses. The CEO of a large-size company that runs a business selling complete home and electronic solutions stated: "Once the businessmen or the ones who run the business in KSA are confident that people have become keen to go online and to visit commerce sites then they will be more than happy to jump on the bandwagon and create their business an e-commerce channel". In order to promote popularity with the customers: "there has to be educational programs". Educational programs make it easy for people to understand how e-Commerce works and what its benefits are. Education is essential in this field and it will necessarily take a while to adopt the culture of e-commerce. Such programs could be started at schools teaching students in their classes about e-commerce. A Regional Director of a medium-size company which runs a business selling home sports equipment said: "It's better to educate students how to use the Internet and benefit well rather than wasting their times on social networks and games!"

These types of programs help people to understand and adapt to the culture of e-commerce and ultimately benefits business by providing a business environment that supports a profitable online market. This addresses the concerns of the owner of a small-size company selling computer related hardware and software who said: "our target, as businessmen, to gain more profit but when there is an idea is not helping us to gain this goal then there is no need to apply or think about it".

This factor was also emphasised by three other participants who have complete e-retailing system solutions in their companies. "The awareness is coming but it takes time and we need to accelerate the development of this field. Like you are conducting this research, this is very good to highlight the difficulties to do something regard them".... "This subject requires education to make it easy for people understand how it works and what its benefits are. After that, when people become well educated, we can benefit from applying such idea like this"....Selling online cannot be compared at all to normal way of selling because the idea of buying online is not adopted by people in Saudi Arabia and "this requires educational programs and building the awareness of e-commerce and other motivated factors".

Educational programs can be formal or informal and can be delivered in many ways. School is just one way. There are online forums, community education programs, TV programs, interest groups, college programs, etc. All of which can be effective at the right time. Education programs are useful for potential shoppers but they are also a requirement for business people to be able to understand a new form of business and to be able to compete on what becomes a non-local marketplace.

## 5.2.    Government Support and Assistance for E-commerce

Most of the participants emphasize the issue of trust of customers for online transactions and this was one area where assistance may be needed from the government. People in Saudi Arabia "have great confidence in anything that comes through the government". "They will have more trust if this subject sponsored by the government". Building trust with customers is a very important issue and: "requires a lot of work and a lot of activity to be there. There are people with 2 to 3 years in e-commerce businesses and selling very well but still very far compared to similar but very small businesses in Europe or Asia... We are working now to have certificates from trusted organizations to build the trust with our customers. Like this there should be a certification body from the government itself to say that this company is a certified company by local government and you can buy from them. This is good to build the customers trust with the certified companies as the government trusts them".

Another assistance needed from the government is to sponsor the solution of e-commerce "taking care of the systems, searching for problems and solutions, and leading toward development". This will encourage sellers to sell online and consumers to purchase online. There should be: "an integrated system such as the Saudi Stock Market. If there is something similar for e-commerce it will be great. Having an integrated system for e-commerce is very important in terms of having people".

In a few words, it is very important for the government to pay high attention on the barriers that retailers face to apply e-retailing systems in order to find solutions and develop initiatives to support the growth of this business type.

## 5.3.    Trustworthy and Secure Online Payment Options

The IT manager of a chocolate and biscuit company which has an e-retailing system said that: "providing more online payments is important because not everybody has a credit card, there should be other ways of online payments". Another participant, the owner of a Saudi manufacturing company for Islamic watches/clocks and his company which has an e-retailing system mainly targeted to international customers, stated "Having more options for online payment other than credit cards will encourage buying online. We currently use in our e-commerce system more than one option for online payment. One option is using PayPal. PayPal is a secure online payment tool for both customer and seller... This system of online payment builds the trust also with the seller who uses it. It represents a company of e-payment and collection that tests new companies that deal with them. For example, they have instructions not to withdraw money for the new sellers before specific period. This procedure is to ensure that the new company is serious and no complaints are received from customers. If there is a complaint, they investigate and return the money to customers if the complaint is true. With this procedure customers feel happy to deal with this intermediary e-payment option which protects their rights and also build the trust with companies that deal with PayPal. So, we think if there is a similar system available in Saudi Arabia and in Arabic language will encourage building trust which is very important factor to buy online".

A regional director of large-sized telecommunication company commented on the unwillingness to use credit cards by people in Saudi Arabia: "I think that this fear is exaggerated. Banks should provide easy options to have two credit cards, one with large amount of money and another one with small amount to be used in online payment. This idea makes people get used to it and this will remove their fear. I suggest for local banks to offer other easy options for online payment to encourage people to purchase online". He continued: "There has to be a mechanism whereby the customers feel confident and will be a 100% trusting to put their credit cards details on the Internet. In order to have that confidence you must have a mechanism whereby fault doesn't happen. And also have secure networking, once we have secure network there will be no problem at all. There are secure payment systems; you just need to incorporate that into your system".

In Saudi Arabia there is an e-payment system called SADAD. SADAD is a national electronic bill presentment and payment service provider for the Kingdom of Saudi Arabia. The core mandate for SADAD is to facilitate and streamline bill payment transactions of end consumers through all channels of the Kingdom's Banks. SADAD was launched on October 3rd, 2004" (SADAD 2004). The head of an Internet services company said "I personally feel more comfortable with SADAD. It is a more secure system and great. With this system there is no need to enter personal payment details on our website. For those who choose to pay using this option, they receive an ID number that is identified in the SADAD system and they use their bank accounts to pay. It is like a bill giving to a customer when paid they can receive their products or services. SADAD is a great idea and more secure than credit cards and encourages people to buy online". However, the IT manager of the chocolate and biscuit company which has an e-retailing system said that SADAD: "before was very expensive solution, it is good for large-size companies but it's not for medium-size company like us!"

## 5.4.    Develop Strong ICT Infrastructure

Infrastructure to provide Internet services that are widely available, reliable and fast is identified as a key facilitator of e-Commerce. The CEO of a large-sized company that runs a business selling complete home and electronic solutions said: "But we need to have the infrastructure right. In order to have the infrastructure right we must make sure that the Wi-Fi is available in better cases and more places. We must ensure that broadband is available for other people in remote places and we must also ensure that the protected plan payment is also taken care of, so that we do not create a system where people actually pay and do not know whether or not the goods will come or not. So it is very important to follow. So getting the infrastructure right is very important".

Mr. Thamer suggested that to get infrastructure right "1) the telecommunication networks inside the city and the country to be at high speed and high efficiency, 2) As much as possible no interruption. If a customer made an online purchase and he/she don't receive the order in the expected time he/she will not be satisfied and trust will be affected. The problem may be not with delivery system, it may be with the system itself not sending/making delay in the order request. 3) Having an integrated system such as Saudi Stock Market. If there is something similar for e-commerce it will be great. Having an integrated system for e-commerce is very important in terms of having people take care of the system, search for problems and solutions, and leading toward development".

## 5.5.    Provision of Sample E-commerce Software to Trial

Most of the participants agree that the provision of an e-commerce software sample to try may be a good motivation to adopt the technology. A participant said "This is a good motivation and easy for seller and buyer". Another participant said "It helps us to understand how it works". Third interviewee commented "this is good and helpful to understanding and help to gain an experience... This experience gives us an idea how it works and how profitable it will be before we get involved".

## 6.    LIMITATIONS AND DIRECTIONS FOR FUTURE RESEARCH
Two limitations need to be acknowledged and addressed regarding the present study. The first limitation concerns lack of access to up-to-date information about retailers, their classification, and statistics. For a number of different reasons, the relevant departments (in government and business sectors) were not helpful in this regard. Callers are referred to official websites where there is no relevant information about retailers' or it is out of date or incomplete. As result, the sample population surveyed in this paper cannot be ensured to be representative of all retailers' types in Saudi Arabia. A second limitation concerns the reluctance of retailers and the difficulty in recruiting participants to complete the interview study. For this reason the selection of participants for interviews was restricted further impacting the representative nature of the sample. The plan was to conduct interviews with 20-30 retailers, however as result of facing this difficulty the number of interviewees is 16.

This study is still in progress. It will be followed by quantitative approach using a survey to test these findings in a wider sample in a range of cities and businesses. This step is very important in order to reduce the limitations in this study, to find the true significance of each of the issues raised, and to verify the completeness of the list of issues determined in the qualitative stage. We will, in due course, be able to report all the factors that positively and negatively affect e-retailing growth in Saudi Arabia and gaining the information from all involved parties in this field in order to contribute to e-commerce development in Saudi Arabia. The collected results from the combination of qualitative and quantitative approaches contribute to the overall research investigating the diffusion of online retailing within the country.

## 7.    IMPLICATIONS AND CONCLUSION
This study's findings will have important practical implications for policy makers in Saudi Arabia in both the public and private sectors. It should assist them to better understand the key factors/areas relating to the diffusion of online retailing and provide a focus in order to guide development. Many of the issues raised in this study have direct parallels to e-commerce development that was undertaken in developed countries in the 1990s. These issues include: creating competitive advantage; expanding to a global marketplace; access to secure online payment gateways; creating trust in an online environment; understanding the vagaries of international trade laws; choosing appropriate goods and services for the online marketplace; creating new business models for the online environment; the need for public education; and, government guidance and support for e-commerce. In those cases there are successful models in literature and in practice that may be emulated where there are similar operating contexts.

Other issues relate to the stage and maturity of nation building and the importance that the nation places on commercial success as a strategic outcome for the success of the country on the international stage. Issues such as the lack of postal addresses; the growth and access to Internet infrastructure; and while relatively immature, are already under constant development as the nation seeks to develop information and communication systems. Relating to the information infrastructure is an identified lack of trained personnel to build and maintain information systems in some government and commercial organizations. With the KSA's move to invest its immense petroleum wealth into creating an information economy for the future there are currently many students undertaking overseas study. There are many international IT providers offering services to fill in the gaps in what can be provided domestically. In the long term the investment in the knowledge economy will mean that there will be a growth in domestic ability to maintain and grow infrastructure as well as export expertise widely.

Saudi Arabia is classified as a wealthy developing nation and as such commercial relationships with the rest of the world, other than for oil sales, are not as important to the governments as for those of many of the western nations. In the western nations it is the commercial advantage of faster computers and Internet access that have driven the success of many companies. There does not seem to be the same developmental drive in the Saudi context. With large national wealth, influence of the Muslim faith, and the power and benevolence of the royal family that drives the different arms of the government, there is a much higher emphasis placed on growing services from the government to the citizen (G2C). In this environment there is not much emphasis being placed on e-commerce at the domestic level and with an existing retail culture there is insufficient competition to drive commercial growth. This is reflected in the difficulties experienced in gathering information for this research project. While government promotion has limited effects on the diffusion of e-commerce in most countries (Kraemer et al. 2006), this study significantly indicates government promotion and support is a key driver to online retailing in KSA.

This study covers the first stage of a much larger project that will cover the perspectives of all stakeholders in the development of e-retail in the KSA. Many of the practical implications and detailed analysis will emerge as the next stages of the study are completed. The findings here will influence the design of the next stage of this research on the retailers' perspectives and a much clearer picture will emerge from the quantitative results.

To conclude, this paper has investigated the issues that positively and negatively influence 16 retailers in Saudi Arabia in the adoption of electronic retailing systems. It reveals a number of perceived impediments relating to cultural, business and technical issues. It also highlights several potential facilitators of retailer adoption of e-retailing practices that will aid development towards e-retail growth in Saudi Arabia. These include access to educational programs and awareness building of e-commerce, government support and assistance for e-commerce, trustworthy and secure online payment options, developing strong ICT infrastructure, and provision of sample e-commerce software to trial. Therefore, policy makers and developers should pay attentions to these factors to facilitate e-retail growth in KSA.

**Appendix**
The Interview questions based on the variables determining the rate of adoption of innovation
of Rogers (2003)

| General Questions | | What are your targeted customers/ marketplace? What do you market/what are your products? What are the most popular goods that you sell? |
|---|---|---|
| | | Does your organization have access to the Internet? Does your company have a website? If so, what kind of information does your organization's website provide? |
| Perceived Attributes of Innovations | **Relative advantage** | Will doing business over the Internet lower your business cost? Is setting up online channel and ongoing maintenance cost cause concern for your organization? Will online channel put your company in better competition position? Do you think the online channel will support traditional retailing? What is your perception about the importance of online channel for your business in the future? |
| | **Compatibility** | What changes needed to apply e-commerce system? Do you think multichannel retailing will be compatible in your company? Has any of your suppliers sell online? If not, would you consider using online channel to sell to your customers cause headache to your company's relationship with the suppliers? Is changing of company's policy and organisational structure are required in order to do business on the Internet? |
| | **Complexity** | Do you find it difficult/confronting to use online system? What makes it difficult advertising your product online? What are the most difficult tasks for your organization selling products online? Does your company have IT department/professionals? Do you have concern about the security of payment over the Internet? Do you have concern that information involved in a transaction over the Internet is not private? |
| | **Trialability** | Have you bought online? If a sample of a system available to try, would you use it? Do you want to try a sample before to apply? Do you see that it is important for Chamber of Commerce/ Ministry of Commerce to/e-commerce solutions companies offer free/cheap deals for trail purposes? |
| | **Observability** | Have you observed online sales activities? When you see those who use the Internet to do business, will this help you to take to decide if your company should go into it as well? How would you see doing business over the Internet generating the desired returns in terms of profit? |

| Communication Channels | What channels do you use to market your products? What are the favourites channels marketing products? How do you most often communicate personally (telephone, in person, e-mail, etc)? |
|---|---|
| **Nature of the Social System** | Do you have any friends in business that do regular online purchasing and selling of goods? What does your social/professional group think about the risks and benefits of e-commerce? If you/when you offer online channel sales, how do you evaluate the reaction of your customers? How would you describe the effects of social system in KSA on online retailing? |
| **Extent of change Agents' promotion efforts** | Are you part of commercial network? Do you receive any business future information/guidelines from Jeddah Chamber of Commerce for example? How do you describe the efforts of chamber of Commerce/ Ministry of Commerce in terms of promoting e-commerce activities in the county? |